\documentclass[article]{mn2e}
\usepackage{epsfig}

\author[Anna L. Watts and Nils Andersson]{Anna L. Watts and Nils Andersson \\
Department of Mathematics, University of Southampton, Southampton, SO17 1BJ,
UK}

\title{The spin evolution of nascent neutron stars}

\begin{document}

\maketitle

\begin{abstract}
The loss of angular momentum due to unstable r-modes in hot young 
neutron stars 
has been proposed as a mechanism for achieving the spin rates inferred 
for young pulsars.  One factor that could have a significant effect on 
the action of the r-mode
instability is fallback of supernova remnant material. The associated 
accretion torque could potentially counteract any gravitational-wave 
induced spin-down, and accretion
heating could affect the viscous damping rates and hence the instability.
We discuss the effects of various external agents on the r-mode instability 
scenario within a simple model of supernova fallback on to a hot young magnetized
neutron star.  We find that the outcome
depends strongly on the strength of the star's magnetic field.  Our model is
capable of generating spin rates for young neutron stars that accord well
with initial spin rates inferred from pulsar observations.  The combined
action of r-mode instability and fallback  appears to cause the spin
rates of neutron stars born with very different spin rates to converge, on a
timescale of about a year.  
The results suggest that stars with magnetic fields $\le10^{13}$~G 
could emit a detectable gravitational wave signal for perhaps several
years after the supernova event.  Stars with higher fields
(magnetars) are unlikely to emit a detectable gravitational wave signal via
the r-mode instability. The model also suggests that the r-mode
instability could be extremely effective in preventing young neutron stars
from going dynamically unstable to the bar-mode.  
\end{abstract}


\section{Motivation}

Neutron stars may, by virtue of their oscillations, act as  sources
of detectable gravitational radiation.  Of particular interest are the r-mode
oscillations, quasi-toroidal currents that arise in rotating stars due to
the action of the Coriolis force.  Andersson (1998) and
Friedman \& Morsink (1998)
first showed that the r-modes of a perfect fluid star succumb to the gravitational radiation
driven Chandrasekhar-Friedman-Schutz instability for all rates of stellar
rotation.  Subsequent results suggest that the instability might grow on short
timescales, with positive feedback increasing the gravitational wave signal
to potentially detectable levels (Owen et al 1998).
The r-mode instability is likely to be most important in hot young rapidly
rotating neutron stars. For such stars there appears to be a window of
temperatures and spin rates in which the r-modes  can grow faster
due to the emission of gravitational waves
than they are damped by viscosity. The consequent
loss of angular momentum provides a plausible mechanism for achieving the spin
rates inferred for young pulsars (Andersson, Kokkotas \& Schutz 1999). In fact, the excitement
regarding the r-mode instability stems to a large extent from the fact that
the first studies suggested that it would spin a newly born neutron star 
to roughly 20~ms, in good agreement with the inferred initial spin period 
of the Crab Pulsar. 

Recent efforts to improve our understanding of the unstable r-modes
have focused on the interior 
physics, aiming to shed light on issues such as the role of general
relativity (Lockitch, Andersson \& Friedman 2001),
superfluidity (Lindblom \& Mendell 2000), the crust/core interface
(Bildsten \& Ushomirsky 2000; Lindblom, Owen \& Ushomirsky 2000), and possible exotic states
of matter like hyperons (Jones 2001; Lindblom \& Owen 2002). As we will show in this paper, 
another factor that may significantly 
affect the r-mode instability is accretion.  
Neutron stars are born in a messy environment, surrounded by debris
from the supernova progenitor. In the first few months some of this material 
will fall back on to the young star.  Intuitively, one would expect
this to affect the action of the r-mode in two ways. Firstly, 
accretion torques could
counteract the gravitational wave induced spin-down, and perhaps 
 cause the star to spin up despite the presence of a large amplitude 
r-mode. One can readily estimate that accretion of 
a total mass $\Delta M$ can spin a neutron star up to
a rotation period
$$
P \approx 3 \left( { \Delta M \over 0.1 M_\odot } \right)^{-1} \hbox{ ms.}
$$
This estimate indicates that rapid supernova fallback may play a key role
in determining the spin evolution of a nascent neutron star.
Secondly, rapid accretion would be associated with strong heating, 
which would affect the strongly temperature dependent 
viscosities,
and hence the longevity and strength of the r-mode
instability and any associated gravitational wave signal.  Accretion
is also expected to delay the formation of a crust until the star has
cooled to a bulk temperature $\sim 10^9$~K (Haensel 1997).  
This should be compared to the case of isolated neutron
stars in which the crust is expected to form at $\sim 10^{10}$~K.
This delayed crust formation could be significant since 
 the presence of a crust would
suppress considerably any unstable r-mode (Bildsten \& Ushomirsky 2000; Andersson \& Kokkotas 2001).  
A delay in crust formation could therefore
increase the longevity of any gravitational wave signal.

Our aim with this paper is to address two
key questions.  Firstly, can the combined action of supernova fallback and r-mode
instability give rise to spin rates that match those inferred for young
pulsars?  Secondly, what are the implications for the 
gravitational wave signal from
the r-mode instability?  In order to address these questions we develop a
simple model of supernova fallback on to a hot young magnetized neutron
star, modelling both spin evolution and gravitational wave emission.

\section{The model}

\subsection{Spin evolution}
In order to be able to compare our results to 
previous work and assess directly the role of the induced 
accretion torque, we employ a simple   
phenomenological spin evolution model analogous to that 
devised by Owen et al (1998) (see also Ho \& Lai 2000; Andersson \&
Kokkotas 2001).  

We assume
that we can model the star's angular momentum as the sum of bulk angular
momentum $J$ and the canonical angular momentum of the r-mode $J_c$.  The total
torque on the star $\dot{J}$ can therefore be written as
\begin{equation}
\label{torque}
\dot{J} = \dot{I}\Omega + I\dot{\Omega} + \dot{J}_c, 
\end{equation}
where the dots indicate time-derivatives. 
$\Omega$ is the star's angular
velocity. For simplicity we will assume that the star does not rotate
differentially despite the presence of a large amplitude r-mode and 
a strong accretion torque. $I =
\tilde{I}MR^2$ represents the moment of inertia of the star. 
We will model the star as an $n=1$ polytrope, which means that 
$\tilde{I} = 0.261$. For this particular stellar model, accretion
will affect the mass $M$  of the star but not the  radius $R$.

In our analysis we will only account for radiation from the 
(dominant) $l=m=2$ current multipole of
the r-mode.  The canonical angular momentum of the mode is
\begin{equation}
J_c = -\frac{3 \Omega \alpha^2 \tilde{J} M R^2}{2},
\label{jcan}
\end{equation}
where $\alpha$ is the mode amplitude (as defined by Owen et al 1998)
and $\tilde{J} = 1.635 \times 10^{-2}$ for an $n=1$ polytrope.

The r-mode is driven by gravitational radiation and damped by viscosity.
We thus assume that the canonical angular momentum,  which
is proportional to the square of the perturbation, evolves according to
\begin{equation}
\frac{\dot{J_c}}{2J_c}= -\frac{1}{\tau},
\label{dotjc}
\end{equation}
where
\begin{equation}
\tau^{-1} = -|\tau_g|^{-1} + \tau_b^{-1} + \tau_s^{-1},
\end{equation}
and $\tau_{g,b,s}$ are the timescales associated with growth of the linear
perturbation due to gravitational radiation 
back reaction and dissipation of energy
due to bulk and shear viscosity, respectively.
From equation (\ref{dotjc}) it can be seen that the
mode is unstable (in an isolated star) 
if $\tau < 0$.  In this case a small perturbation will
lead to an unbounded growth of the mode.  

Equations (\ref{torque}), (\ref{jcan}) and (\ref{dotjc}) can be combined 
to give
equations for the evolution of $\alpha$ and $\Omega$:
\begin{equation}
\label{dotalphaacc}
\dot{\alpha} = - \alpha\left[\frac{1}{\tau}+\frac{\dot{\Omega}}{2\Omega}+\frac{\dot{M}}{2M}\right],
\end{equation}
and
\begin{equation}
\label{dotomegaacc}
\dot{\Omega} =
\left[\frac{\dot{J}}{I}-\frac{\dot{M}\Omega}{M} -\frac{3 \tilde{J} 
\alpha^2\Omega}{\tilde{I}\tau}\right].
\end{equation}
We initiate our evolutions by assuming that a small amplitude r-mode
(with $\alpha= 10^{-6}$) is present. Variation of the initial amplitude has
no significant effect on the subsequent  evolution. 

Equations (\ref{dotalphaacc}) and (\ref{dotomegaacc}) only apply when
$\alpha$ is small.  To model the behaviour once the r-mode
reaches the non-linear regime we assume that 
it saturates at some value $\alpha_s$.  The larger
the value of $\alpha_s$, the greater the fraction of the star's angular
momentum the r-mode can carry (we consider $0.01\le\alpha_s\le 1$ in this
paper).   When the mode is saturated the spin
evolution is described by 

\begin{equation}
\dot{\Omega} = \frac{\dot{J}}{I}\left[1-
\frac{3\tilde{J}\alpha_s^2}{2\tilde{I}}\right]^{-1}  -
\frac{\dot{M}\Omega}{M}.
\label{dotomegasat}
\end{equation}
As discussed by Owen et al (1998), the bulk of gravitational wave emission and
 associated spin-down will occur
during the saturated phase.

Eventually the mode
will become stable again, de-saturate and die away; when this happens we
switch back to using equations (\ref{dotalphaacc}) and
(\ref{dotomegaacc}). Note that equation (\ref{dotalphaacc}) indicates that the r-mode is unstable, 
in the sense that $\dot{\alpha}>0$, 
whenever
\begin{equation}
\label{inst}
\frac{1}{\tau}+\frac{\dot{\Omega}}{2\Omega}+\frac{\dot{M}}{2M} < 0.
\end{equation}
This is the appropriate criterion for the onset of instability in a star that
is spun up by accretion (Ho
\& Lai 2000).  Compare this with the instability criterion for
an isolated star, $\tau < 0$. 

We use the estimates given by Andersson \& Kokkotas (2001) for the 
gravitational radiation reaction timescale due to the  $l=m=2$ current 
multipole:  
\begin{equation}
\tau_g \approx -47 M_{1.4}^{-1} R_{10}^{-4} P_{-3}^6 \mathrm{~s,}
\end{equation}
as well as the bulk and shear viscosity damping timescales:
\begin{equation}
\label{bulk}
\tau_b \approx 2.7\times 10^{11} M_{1.4} R_{10}^{-1} P_{-3}^2 T_9^{-6} \mathrm{~s,}
\end{equation}
\begin{equation}
\tau_s = 6.7 \times 10^7 M_{1.4}^{-5/4} R_{10}^{23/4} T_9^2 \mathrm{~s,}
\end{equation}
where we have used the notation $M_{1.4} = M/1.4M_\odot$, $R_{10} =
R/10$~km, $P_{-3} = P/1$~ms and $T_9 = T/10^9$~K.  We have used these
timescales because it allows us to make a direct
comparison with the earlier work of Owen et al (1998).  However, it should
be noted that this is
probably the most optimistic scenario with regard to the lifetime of
unstable r-modes.  We do not take into account any damping effects that may
arise from the interaction of mode and magnetic field (Rezzolla, Lamb \& Shapiro
2000), or from the presence of
exotic particles such as hyperons (Jones 2001: Lindblom \& Owen 2002).

\subsection{Torques}

The total torque on the star $\dot{J}$ is taken to be the sum of torques due to
gravitational radiation, accretion and magnetic dipole radiation.  The
torque due to gravitational-wave emission from the $l=m=2$ current
multipole is 
\begin{equation}
\dot{J}_g = 3 \tilde{J} \Omega \alpha^2 M R^2 \tau_{g}^{-1}. 
\end{equation}
In order to model the accretion torque we need to know the quantity and
 rate of fallback onto the young neutron star. Several authors have
 examined this problem and concluded that a significant quantity of matter
 (up to $\sim 0.1M_\odot$) could fall back onto the neutron star during the
 first few months of its life (Colgate 1971; Chevalier 1989; Houck \&
 Chevalier 1991). Accretion of such a
 large amount of matter onto a neutron star is hypercritical;
 that is, orders of magnitude above the Eddington limit (compare with
 Bethe, Brown \& Lee 2000). 
 Such rapid accretion would be advection dominated, with energy loss
 occurring via neutrino emission as matter is processed at the surface of
 the neutron star.  Advection dominated accretion flows have been examined
 in detail by Narayan \& Yi (1994,1995).  They conclude that
 advection could dominate in cases where the mass accretion rate is very
 high; precisely the conditions likely to prevail around a newly-formed
 neutron star. 

Advection dominated accretion onto a magnetized neutron star has been
examined by Menou et al (1999).  Building on earlier work (for example
Illarionov \& Sunyaev 1975; Ghosh \& Lamb 1978), they model a rotating magnetized
neutron star surrounded by a magnetically threaded accretion disk.
Accreting matter follows magnetic field lines and gives up angular momentum
on reaching the surface, exerting a spin-up torque.  The contribution of
the magnetically threaded disk is however more complex.  Interaction
between the stellar field and the disk results in a positive torque for
small radii, where the field lines rotate more slowly than the local
Keplerian speed of the gas.  For larger radii, the field lines rotate more
quickly than the local Keplerian speed, resulting in a negative torque.  If
the spin period becomes very short, or the rate of fallback of material
onto the magnetosphere drops, the star can enter what is known as the
propeller regime. Then the rapidly rotating magnetosphere exerts a
centrifugal barrier that inhibits 
further accretion.  Accreting matter is flung away
from the star and the star experiences a spin-down torque.
An accreting
magnetized star can therefore experience either a spin-up or a spin-down
torque.

To proceed we need to determine the likely rate and quantity of supernova
fallback.  Mineshige et al (1997) model the supernova fallback problem assuming that the fallback material possesses some angular momentum (compare with Chevalier (1989) and Brown \& Weingartner (1994), who  model the
 problem under the assumption of Bondi spherical
accretion).  Mineshige et al find a fallback rate 
$\dot{M}_f \propto t^{-n}$, with $1<n<2$.   We therefore use
\begin{equation}
\label{accrate}
\dot{M}_f = Ct^{-3/2},
\end{equation}
where $C$ is set by choosing the total mass to be accreted $M_{acc}$
between some initial time $t_i$ after the supernova explosion and
$t=\infty$.  The accretion rate is related to the 
fallback rate by
\begin{eqnarray}
\dot{M} = \left \{ \begin{array}{ll} \dot{M}_f & \mathrm{Propeller~off,} \\ 0 &
\mathrm{Propeller~on.} \end{array}
\right.
\end{eqnarray}
Following Chevalier (1989) and Mineshige et al (1997) we choose $M_{acc}$ to be $\le
0.1M_\odot$.  We choose $t_i=100$~s as this is the time after the supernova
at which we expect there to be a recognisable compact object (Burrows \&
Lattimer 1986).
If the propeller effect is not active, we assume that hypercritical
accretion continues until the accretion rate falls to $\dot{M}
\le 10^{-4}M_\odot \mathrm{yr}^{-1}$.  This 
corresponds to the cut-off point below which hypercritical accretion is 
no longer possible. 
 We follow Chevalier (1989) and Bethe et al (2000) in assuming that
below this rate photons begin to diffuse out of the shocked region
and further accretion proceeds at (or below)
the Eddington rate ($\sim 10^{-8}M_\odot \mathrm{yr}^{-1}$).  

If on the
other hand the propeller effect becomes active, the situation is less
clear.  The propeller effect prevents accretion and trapped photons 
may escape as soon as the propeller becomes active.  But would the
pressure of these photons reduce the rate of fallback of material on to the
magnetosphere?  We model two possibilities: i) continued exponentially
decaying fallback, and ii) constant accretion at the Eddington rate.  
For stars
with magnetic fields of $\le10^{13}$~G there is little quantitative
difference between the two models, however there are noticeable differences
for higher fields.

A typical young neutron
star is expected to have a magnetic field $B\sim 10^{12}$~G (see Table
\ref{pulsars}), although a magnetar could have
a field strength as high as $10^{15}$~G (see for example Thompson \&
Duncan 1996).  We
assume that the magnetic field strength is not affected by accretion or the
action of the mode, 
even though it is by no means clear that this
would be the case in practice.  Several authors 
have suggested that the
r-mode might lead to differential rotation and hence to changes in the
magnetic field (see for example Spruit 1999 and 
Rezzolla et al 2000).
In the earliest stages after the supernova the accretion rate is likely to
be so high that the magnetic field of the neutron star is completely
confined. The pressure of accreting material will be so much greater than
the magnetic pressure that the magnetospheric radius, 
defined by
\begin{equation}
r_m = \left[\frac{B^2 R^6}{2\dot{M}(2GM)^{1/2}}\right]^{2/7},
\end{equation}
with $B$ taken to be the polar magnetic field of the star, 
is smaller than the radius of the neutron star.  The
magnetic field remains confined to the neutron star until the accretion
rate falls sufficiently for the field to begin to exert an influence
outside the star.

The accretion torque $\dot{J}_a$ will vary depending on whether the
magnetic field is confined and whether the propeller effect is
operational.  When the field is completely confined we use Narayan and Yi's (1994,1995) results
for the accretion torque.  Their models of advection
dominated accretion flows suggest orbital velocities with $v^2 \approx
\frac{2}{7}v^2_k$, $v_k$ being the Kepler velocity of a particle
orbiting the star. Hence
\begin{equation}
\dot{J}_a = \dot{M} R \left[\frac{2GM}{7R}\right]^{1/2}.
\end{equation}
As the accretion rate falls, $r_m$ grows and eventually exceeds the neutron
star radius $R$.  At this stage we switch to the equations of Menou et al
(1999), as the torque will now be affected by the interaction of the
magnetic field and the fallback material.
\begin{eqnarray}
\label{ja}
\dot{J}_a = 2  r_{m}^2 \Omega_{k}(r_{m})\left[1-\frac{\Omega}{\Omega_{k}(r_{m})}\right] \times
\left \{ \begin{array}{ll}  \dot{M}
 & \mathrm{Propeller~off,} \\ \dot{M}_f & \mathrm{Propeller~on,}
\end{array}
\right.
\end{eqnarray}
where $\Omega_k(r_m)$ is the Keplerian angular velocity evaluated at the
magnetospheric radius.  This magnetized accretion torque falls as the star
spins up, and for $\Omega > \Omega_k(r_m)$ the torque will in fact be negative.
The star is then in the propeller regime.  
During this phase matter is prevented
from accreting onto the star and fallback material is instead expelled from
the system by the rotating magnetosphere.  Menou et al (1999) review the
assumption that accretion ceases and conclude that accretion is indeed
minimal when the propeller is active. 
We do not attempt to model the complicated transition regime 
between the two phases in detail.

Once magnetic field confinement ends
we could also include a torque due to emission of magnetic dipole radiation,
$\dot{J}_m$, where
\begin{equation}
\dot{J}_m = -\frac{2 B^2 R^6 \Omega^3}{3 c^3}.
\end{equation}
This term was however found to be negligible compared to the other torques
acting on the young neutron star. Its effects will not be discussed
further in this paper.

\subsection{Temperature evolution}

Viscosity, the main agent damping the r-mode in our model, 
depends critically on
temperature.  We include three factors in modelling the temperature
evolution: modified URCA cooling, shear viscosity reheating, and accretion
heating.  We assume the star to be isothermal even though
this may not be appropriate
for a newly born accreting star (Burrows \& Lattimer 1986). Still, this 
assumption  simplifies the analysis 
considerably and all previous studies of the r-mode 
instability have assumed a uniform temperature distribution.
 
The primary cooling mechanism for a young neutron star will be the modified
URCA reaction, 
in which neutrons decay via the weak interaction, emitting
neutrinos.  We do not model any rapid cooling effects due to the direct 
URCA reactions in the core of the star (compare with Lattimer et al 1994 and
Page et al 2000). 
The cooling rate due to the modified URCA reaction,
$\dot{\varepsilon}_u$,
 is given by Shapiro \& Teukolsky (1983):
\begin{equation}
\label{urcacool}
\dot{\varepsilon}_u= 7.5\times10^{39} M_{1.4}^{2/3} 
T_9^8 \mathrm{~erg~s}^{-1}.
\end{equation}
The neutron star will be heated by the action of shear viscosity on the
r-mode oscillations.  The
heating rate due to shear viscosity, $\dot{\varepsilon_{s}}$, is given by
Andersson \& Kokkotas (2001):
\begin{eqnarray}
\label{shearheat}
\dot{\varepsilon}_s & = & \frac{2 \alpha^2 \Omega^2 M R^2
\tilde{J}}{\tau_s}{} \nonumber \\ &=& 8.3\times 10^{37} \alpha^2 \Omega^2
\tilde{J} M_{1.4}^{9/4}
R_{10}^{-15/4}
T_9^{-2} \mathrm{~erg~s}^{-1}.
\end{eqnarray}
Accretion heating will have two components.  The first contribution arises
when accreting matter undergoes nuclear burning at
the surface of the star.  We assume that
accreting matter will release approximately 1.5 MeV of energy per
nucleon (Brown \& Bildsten 1998).  If we assume that every nucleon reaching the
surface is burnt, then energy is 
liberated at a rate:
\begin{eqnarray}
\dot{\varepsilon}_n & = & \frac{\dot{M}}{m_B} \times 1.5~\mathrm{MeV}{}
\nonumber \\ &=& 4\times 10^{51} \dot{M}_{1.4} \mathrm{~erg~s}^{-1}
\end{eqnarray}
where $m_B$ is the mass of a baryon.  

A second contribution arises because the flow is assumed to be advection
dominated.  Matter falling in towards a star will liberate $\sim GM/R$ 
of potential energy per unit mass.  In a non-advection
dominated flow 	most of this energy
would be dissipated as heat in the accretion disk.  In
an advection dominated flow, by contrast, this energy is carried in with
the flow of matter. We assume that the advected potential energy is used to
generate neutrinos near the surface of the star, 
and that these neutrinos are radiated isotropically from
the point of generation.  We therefore assume that half of the neutrinos escape without
interacting with the star and that the remaining half are radiated into the
star, where they scatter and interact with the stellar material.  The mean
free path $\lambda$ of inelastic scattering of neutrinos with electrons in
the neutron star matter is
\begin{equation}
\lambda \sim 2\times10^{10}\left[\frac{\rho_{nuc}}{\bar{\rho}}\right]^{7/6} 
\left[\frac{0.1~\mathrm{MeV}}{E_{\nu}}\right]^{5/2}~\mathrm{cm,} 
\end{equation}
where $\rho_{nuc} \sim2.8\times10^{14} ~\mathrm{g~cm}^{-3}$ and
$\bar{\rho}$ is the average density of the star. For an $n=1$
polytrope we have
\begin{equation}
\frac{\bar{\rho}}{\rho_{nuc}}= 0.7\frac{M}{M_\odot}.
\end{equation}  
We assume
that the neutrino energy is $E_{\nu}= 1$~MeV (Shapiro \& Teukolsky 1983).  Assuming fully
efficient scattering, and noting that $\lambda>R$, the heating rate due to advected potential energy may
be estimated as
\begin{eqnarray}
\dot{\varepsilon}_h  & \sim &
\frac{R}{\lambda}\frac{GM\dot{M}}{R}{} \nonumber \\ &=& 8\times10^{51}
M_{1.4}^{13/6} \dot{M}_{1.4} \mathrm{~erg~s}^{-1}.
\end{eqnarray}
We estimate the heat capacity $C_{\nu}$ by assuming the thermal energy of
the neutron star to reside almost exclusively in degenerate neutrons
(Shapiro \& Teukolsky 1983).  Neglecting interactions, the heat capacity of such a system
is given by
\begin{equation} 
\label{heatcap}
C_{\nu}= 1.6\times10^{39} M_{1.4}^{1/3} T_9 \mathrm{~erg ~K}^{-1}.
\end{equation}
Equation (\ref{heatcap}) shows that $C_{\nu}\propto T$, hence the equation of thermal balance of the star is
\begin{equation}
\label{thermalbalance}
\frac{d}{dt} \left[\frac{1}{2} C_{\nu} T\right] = 
\dot{\varepsilon}_s - \dot{\varepsilon}_u + \dot{\varepsilon}_n + \dot{\varepsilon}_h.
\end{equation} 
Manipulation of equation (\ref{thermalbalance})
yields an expression for the temperature evolution of the star. We initiate
our evolutions at $T=10^{11}$~K and
terminate them at $T=10^9$~K, the temperature at which we expect
a crust to form (Haensel 1997) and suppress the r-mode instability 
in an accreting star.

\section{Observational constraints}

Before we proceed to the discussion of the results obtained from 
our simple model it is useful to consider the constraints 
provided by observations of young neutron stars. 
It is standard practice to assume that a pulsar spins
down purely as a result of magnetic dipole radiation.  
Given the observed period and the spin-down rate we can then 
 estimate the magnetic field
strength from
\begin{equation}
B = 3.2\times10^{19}(P \dot{P})^{1/2} ~\mathrm{G.}
\end{equation}
Similarly, one can infer 
the characteristic age $t_c$ of the
pulsar from the observed data:
\begin{equation}
t_c= \frac{P}{2\dot{P}}.
\end{equation}
Here it is assumed that the star was born spinning
much faster that its current rate. 
The use of this estimate is supported by the fact that, in the 
case of the Crab pulsar, the
age calculated in this way accords well with the known
age. 

If observations also provide the braking index, $n$, and an estimate of
pulsar age, $t_{SNR}$,
the initial spin period of the pulsar $P_0$
 is inferred from
\begin{equation}
P_0 = P \left[ 1 - (n-1) t_{SNR} { \dot{P} \over P }\right] ^{1/(n-1)},
\end{equation}
where $n$ is given by 
\begin{equation}
n = \frac{P\ddot{P}}{\dot{P}^2} - 2.
\end{equation}

In most cases the largest uncertainty in estimating the 
initial spin period is associated with the age
of the pulsar, $t_{SNR}$. Ideally, one would like to make an
association with a known historical supernova. This would then, 
as in the case of the Crab, provide a precise age estimate 
and a reasonably reliable initial spin period. Unfortunately, 
this kind of data is the exception rather than the norm. 
Typically, the independent age is estimated by measuring
the size of the supernova remnant and comparing it with theoretical models
for the expansion rate. This procedure
is complicated by many factors and the results are
associated with large error bars.  

A sample of data for young pulsars that are claimed to be associated with
known supernova remnants is given in Table~\ref{pulsars}.
The reliability of the inferred initial spin periods
can be assessed by comparing the characteristic age of the 
pulsar to the estimated age of the supernova remnant. 
For most of the cases listed in Table~\ref{pulsars},  
these estimates are consistent.  In these cases the association between
supernova remnant and pulsar seems correct and the spin of these pulsars
appears to have evolved primarily due to the action of magnetic dipole
radiation. 

The data for young pulsars clearly support the notion that 
neutron stars are not formed spinning near the Kepler 
limit. This would be consistent with both the r-mode 
instability scenario (where a rapidly spinning, 
newly born neutron star is 
spun down to a period of say 10~ms in a few months) 
and the possibility that magnetic core-envelope 
coupling in the supernova progenitor leads to 
most neutron stars being born rotating slowly (Spruit \& Phinney 1998). 

There is ongoing discussion in the literature as to whether
the (obviously simplistic) magnetic dipole model adequately describes the spin
evolution of young pulsars.  Concerns stem from discrepancies between
various age estimates for some young pulsars.  Calculations would also be
complicated by magnetic field decay (Colpi et al 2000) or
growth (Gaensler \& Frail 2000). Another problem is that all
measurements of the braking index $n$ are lower 
than the value of 3 predicted by the magnetic dipole model. Several authors
have suggested that the propeller effect, driven by ongoing low rate
accretion of supernova remnant material, could explain the discrepancies
(Marsden, Lingenfelter \& Rothschild 2001; Menou, Perna \&
Hernquist 2001). The same mechanism might also explain the properties of
the Anomalous X-Ray Pulsars (AXPs) (Perna, Hernquist \& Narayan 2000;
Chatterjee, Hernquist \& Narayan 2000; Alpar 2001; Menou, Perna \& Hernquist 2001),
without having to invoke large magnetic field strengths (Thompson \& Duncan
1996). See Rothschild, Lingenfelter \& Marsden (2001) for a review of AXP properties
and the competing accretion-driven and magnetar models.

The relation of these propeller/fallback models to the one used in this
paper is simple.  Because we are interested in gravitational wave emission,
we model only the very earliest stages of fallback and terminate our
evolutions at crust formation.  At crust formation ($t\sim 1$~year) there is still
material in the fallback disk, although the fallback rate has dropped
significantly.  The papers listed above focus more on the subsequent long
period of sub-Eddington accretion.  Although they do discuss the early
super-Eddington accretion phase (see for example Chatterjee, Hernquist \&
Narayan 2000), they do not model this phase in detail and neglect the
effects of gravitational wave emission completely.

\section{Results from the fallback model} 

We are now prepared to discuss the results obtained 
from our phenomenological r-mode instability/supernova
fallback mode, and compare the predicted spin rates 
to the data in Table~\ref{pulsars}.

If we consider the data in Table~\ref{pulsars}, and the 
fact that the r-modes are certainly stable for periods longer
than (say) 20~ms (see Andersson \& Kokkotas 2001), it would seem as if the instability 
may only be relevant for a small sample of nascent
neutron stars. This would be in agreement with 
Spruit \& Phinney's (1998) model, which suggests that most 
neutron stars are born rotating slowly. 
Our study will show, however, that
supernova fallback may change this picture considerably, with accretion induced
spin-up driving the star into the window where the r-mode is likely
to be unstable. In other words, significant supernova fallback  
enhances the probability that a young neutron star evolves
through a phase where the r-mode instability is active.  

\begin{figure}
\hbox to \hsize{\hfill \epsfysize=6cm
\epsffile{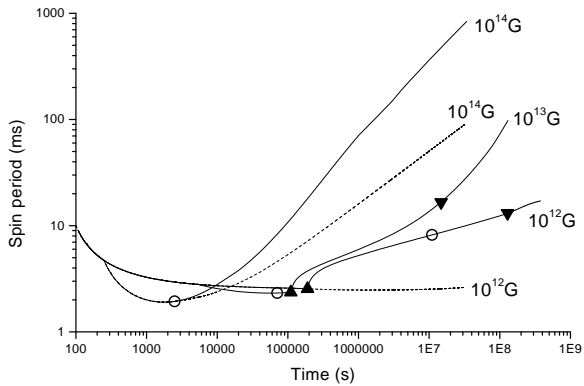} \hfill}
\caption{The solid lines show the evolution of the spin period for newly-born accreting neutron stars
with an unstable r-mode and
 different magnetic field strengths [Key:  Hollow circles - propeller
switches on; Upwards triangles - r-mode saturates; Downwards triangles -
r-mode becomes stable].  Note that the r-mode never reaches the saturation amplitude 
for the star with the highest magnetic field.  The dashed lines, given for
comparison for the two extreme magnetic field strengths, show the spin
evolution if an unstable r-mode is not present.
}
\label{magp}
\end{figure}

Let us consider the effect of supernova fallback on a
neutron star born spinning at a reasonably small
fraction of the Kepler rate. Taking the birth spin period as 10~ms, 
we evolve the star according to our phenomenological model. 
Typical results for the resultant spin evolution 
are shown in Figure~\ref{magp}.  The data shown in the 
figure suggest the following picture. After an initial spin-up 
phase, the r-mode becomes unstable and begins to grow.  For a typical
young neutron star with magnetic field
$\sim10^{12}$~G the r-mode instability 
provides the main spin-down mechanism.
For larger magnetic fields the propeller effect becomes active at an earlier
stage of the evolution and  
is the main spin-down mechanism for highly magnetized stars.  Nevertheless,
the r-mode is important in even the most highly
magnetized stars.  Switching off the r-mode leads to a marked reduction in
spin-down, despite the fact that the mode never actually saturates in the star with
the largest magnetic field.  Note also that our model
predicts that supernova fallback will
delay the appearance of the r-mode signal compared to 
the case of an isolated star, for which the r-mode is predicted to saturate
after $\sim 10$~minutes (Owen et al 1998).  This can be seen from equation
(\ref{inst}): the high initial accretion rate overwhelms all of the other
terms in the expression and $\alpha$ will not grow until $\dot{M}$ has
fallen.

In our model we have assumed that the magnetic field does not
influence $\dot{M}_f$
(although it does affect $\dot{M}$ via the propeller effect).  Recent
work by Igumenshchev \& Narayan (2002) on spherical fallback, however,
suggests that the presence of a magnetic field may reduce $\dot{M}_f$ by
perhaps a factor of 10. If we assume such a reduced $\dot{M}_f$
from the time at which magnetic field suppression ends, we find little
impact on spin rate or gravitational wave emission for a $10^{12}~G$ star.
The effect on a $10^{14}$~G star is however more pronounced. Initial
spin-up and subsequent propeller-driven
spin-down are much reduced, as is the emitted gravitational wave amplitude. Note
however that we only examine the first $\sim 1$~year
post-supernova and that the effects of slower fallback may be more
important at later times.

\begin{figure}
\hbox to \hsize{\hfill \epsfysize=6cm
\epsffile{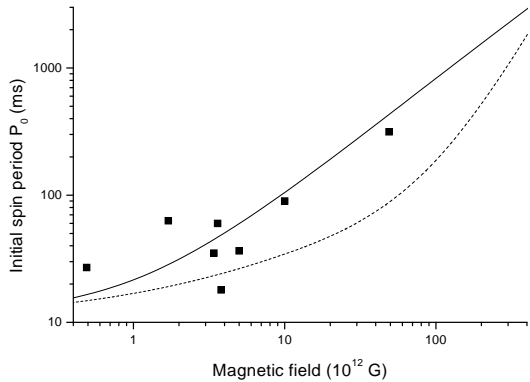} \hfill}
\caption{Spin rates predicted by the fallback model at crust formation
(within 1-10 years of the supernova) compared to values inferred for young
pulsars using the standard magnetic dipole spin-down model (black squares).
The solid line is for a model where fallback rate continues to decay
exponentially when the propeller switches on; the dashed line for one where
the fallback rate then drops to the Eddington limit. [Note that the 
observational data is associated with large error bars which are not included in the figure.]}
\label{compare}
\end{figure}

We have terminated the evolutions once the star reaches a 
temperature of $10^9$~K. At this point one would expect the crust 
to form, which is predicted to increase dramatically viscous 
damping of the r-mode (Bildsten \& Ushomirsky 2000; Lindblom et al 2000). 
In  Figure~\ref{compare} we compare the spin period
at the end of the our evolutions to the
values of $P_0$ inferred for young pulsars using the magnetic dipole model
(as listed in Table~\ref{pulsars}).
This figure should, of course, be considered with caution
since the inferred spin periods are associated with significant error bars.
Still, our model clearly captures the main trend of the data,   
namely that the initial spin period inferred from the magnetic dipole model increases with
magnetic field. Given the many uncertainties in both 
the theoretical model and the observational data the agreement 
between the results is rather good.

\begin{figure}
\hbox to \hsize{\hfill \epsfysize=6cm
\epsffile{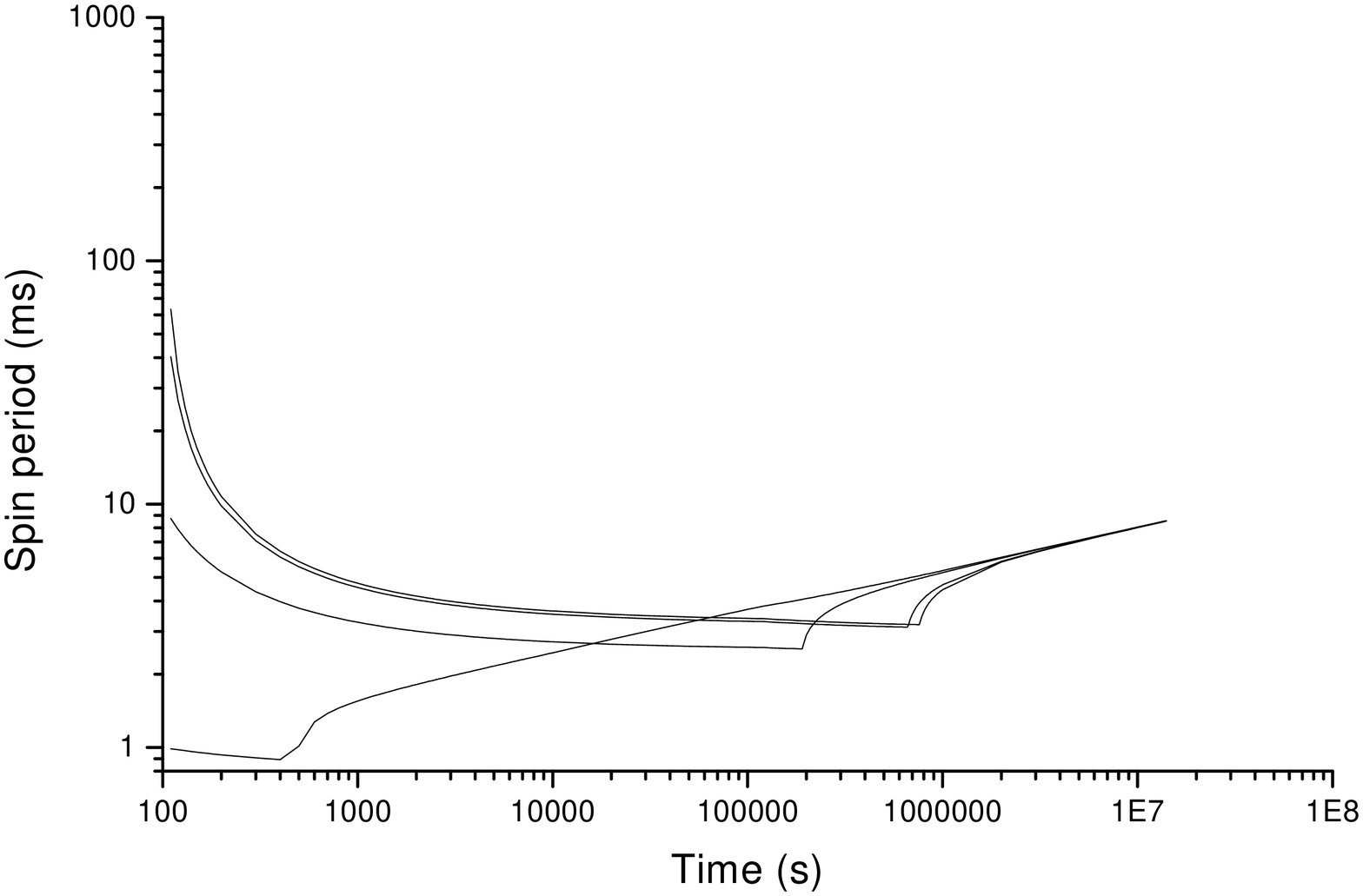} \hfill}
\caption{The combined action of the r-mode instability and fallback
accretion causes the spin rates of stars with birth spin
rates in the range 1~ms to 1~s to converge to $\sim 10$~ms within $\sim 1$
year.  The data shown are for a star with $B=10^{12}$~G.}
\label{pip}
\end{figure}

The fallback model suggests that the outcome is insensitive to 
the spin period at the beginning of the evolution (the true
``birth'' spin rate).  
The combined action of supernova fallback
and the r-mode instability causes the spin rates of stars with very
different birth spins to converge within $\sim 1$ year.  This is shown in
Figure \ref{pip}.  

This result is notable given the uncertainty over
the spin rates at which neutron stars are formed.  It may, in fact, 
suggest that the true ``birth'' spin is not the key factor that 
determines the spin period at (say) one year post-supernova. 
This observation has  implications for determining 
whether or not a young neutron star
undergoes a phase where the r-modes are unstable. This will in turn
affect the relevance of the r-modes as a gravitational wave
source. Of particular interest is the fact that our results suggest
that fallback accretion may lead to even a very slowly spinning
neutron star being spun up sufficiently for the r-modes 
to become unstable after $\sim10$~min. However, we conclude that the
magnetic field is far more important in determining the
spin rate at one year post-supernova than both the birth spin rate
and the possible r-mode spin-down phase.

\begin{figure}
\hbox to \hsize{\hfill \epsfysize=6cm
\epsffile{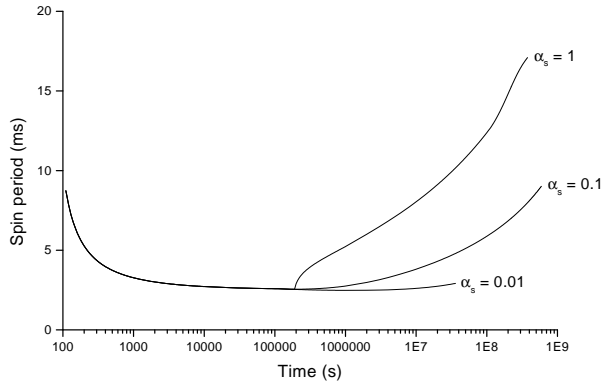} \hfill}
\caption{The effect on spin evolution of changing $\alpha_s$ for a star with $B=10^{12}$~G.}
\label{alphasp}
\end{figure}

The key parameter in determining the role of the r-mode instability
is the saturation amplitude $\alpha_s$. As has been demonstrated 
by recent hydrodynamical simulations (Stergioulas \& Font 2001, 
Lindblom, Tohline \& Vallisneri 2001), 
 the 
r-modes may not saturate until the amplitude has reached
values of order unity.  
The effect of varying $\alpha_s$ is shown 
in Figure \ref{alphasp}.  These results are exactly what one would expect: 
the r-mode needs to be able to grow to a large amplitude if it is to
counteract the strong accretion torque during the fallback phase.

\begin{figure}
\hbox to \hsize{\hfill \epsfysize=6cm
\epsffile{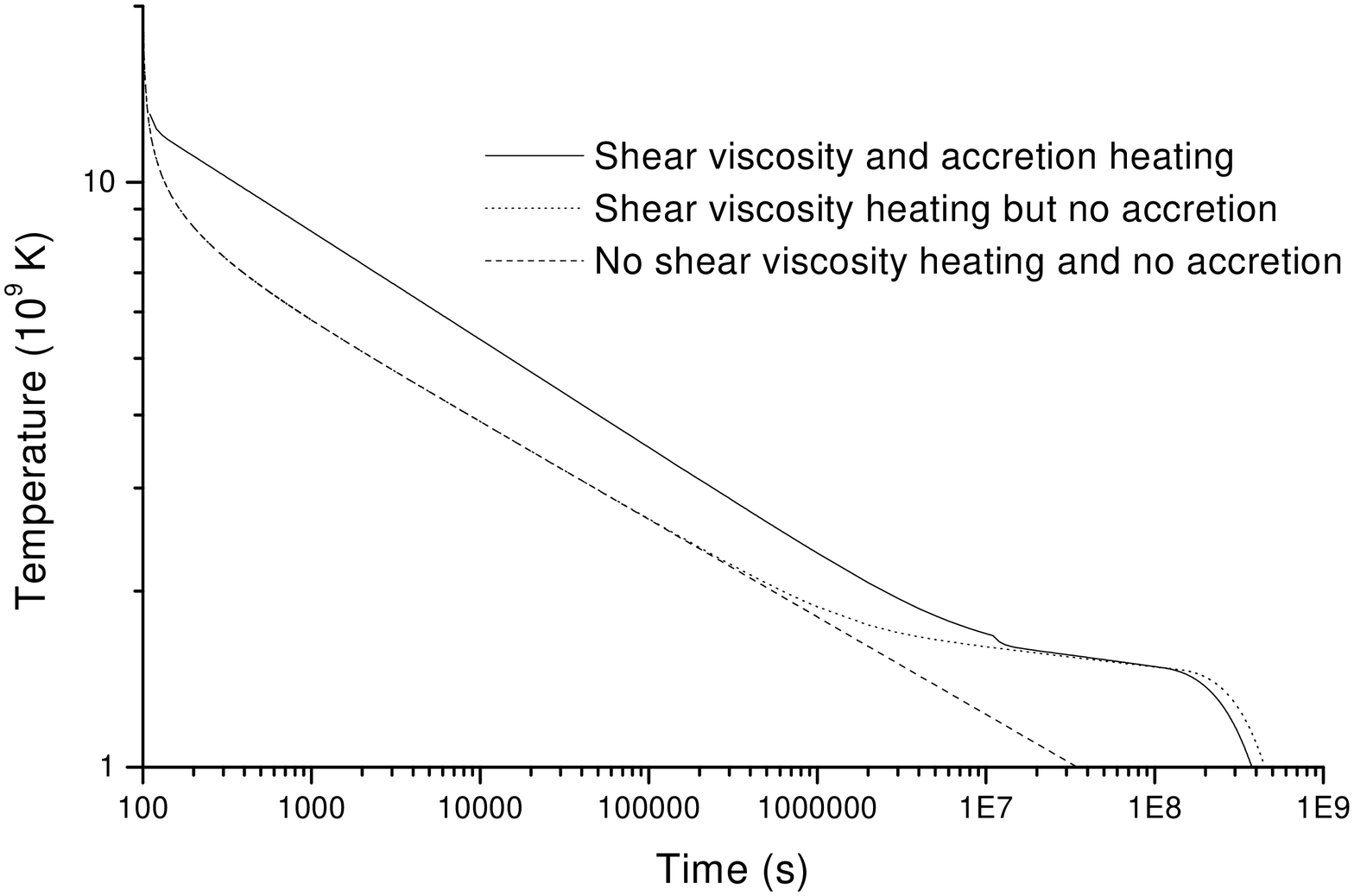} \hfill}
\caption{The effect of shear viscosity reheating and accretion heating on
the temperature evolution of a young neutron star.}
\label{temp}
\end{figure}

One might expect fallback accretion to lead to 
strong heating, and that this would have a significant effect 
on the duration of the r-mode gravitational-wave signal. 
We analyse the role of accretion heating in Figure~\ref{temp}.
We find that even though rapid accretion increases the temperature
by roughly 10 percent throughout the initial evolution, it does not 
lengthen the r-mode instability phase.  
Far more important in this respect is 
reheating due to the action of shear viscosity on the r-mode, which
prolongs the r-mode instability phase for both an accreting and a
non-accreting star.  The amount of viscous heating is
essentially the same in our current model as it would be in the case of an
isolated young neutron star. 

In order to assess the detectability of the gravitational 
waves due to the r-modes, we use the prescription of 
Owen et al (1998). The true strain of the gravitational waves, $h$, is
given by 
\begin{equation}
\label{h}
h(t) = 7.54\times10^{-23} \alpha \tilde{J} \Omega^3 M_{1.4} R_{10}^3
\frac{15~\mathrm{Mpc}}{D}.
\end{equation}
We consider sources in the Virgo Cluster, for which the distance to the
source, $D\approx 15$~Mpc.  Results from our model are similar to results
from previous studies (such as Owen et al 1998).  They show that the unstable r-modes
may lead to a gravitational wave signal that could 
be detectable
by a second generation detector like LIGO II. 
The main difference between our results and those of Owen et al (1998)
is the duration of the signal. We predict that, 
if continued accretion were to take place and prevent 
crust formation until the core temperature has fallen to 
$10^9$~K, the signal may last for several years. We find that the peak gravitational-wave signal
amplitude remains almost unchanged 
as we vary the saturation amplitude by two orders of
magnitude, even though this has a noticeable effect on 
final spin period. 
Note however that these conclusions are only relevant for neutron stars with
fields $<10^{13}$~G.  It is unlikely that we would observe
a signal from the r-modes of neutron stars with larger magnetic fields.
This conclusion agrees well with studies of Rezzolla et al (2000)
and Ho \& Lai (2000). In our case the 
reduction in signal with increasing magnetic field is 
due mainly to increased propeller effect spin-down.  This is because the 
gravitational wave amplitude depends so strongly on spin rate (see equation
(\ref{h})) .  It should of course be pointed out that we have not taken account 
of any possible interaction between the
r-mode and the magnetic field, which may further decrease
the relevance of the r-modes in magnetized stars.

Finally, we have considered the possibility
that the fallback accretion torque
might  be able to spin the star up to (and beyond) the mass-shedding limit
(for the canonical $n=1$ polytropic neutron star used in our evolutions
this corresponds to $P \approx 0.8$~ms).
Depending on the overall stiffness of the supranuclear equation of state, 
neutron stars spinning near the mass-shedding limit may be subject to 
the dynamical bar-mode instability (Shapiro \& Teukolsky 1983). This would lead to 
the formation of a highly non-axisymmetric bar-like
configuration which is expected to be a strong source 
of gravitational waves with
frequencies of order kHz (Houser, Centrella \& Smith 1994).  Whether a young neutron star
becomes bar-mode unstable is therefore of great interest to gravitational
wave theorists. Interestingly, our evolutions indicate that the r-mode
instability is remarkably efficient in preventing the star 
from reaching the mass-shedding limit during the fallback phase.
Even stars born with periods as short as 0.85 ms were unable to reach
the Kepler threshold.  The r-mode grows rapidly and is able to spin down the
star, even in the initial stages where the accretion torque is the highest.
This suggests that r-mode induced spin-down might counteract accretion
spin-up to prevent all but the most rapidly rotating young neutron stars
from becoming bar-mode unstable.  For comparison we also investigated the case where there
was no unstable r-mode.  In this case the propeller effect switches on far
earlier and can also act to prevent the star from going bar-mode unstable.
Consider the $B=10^{12}$~G star shown in Figure~\ref{magp}, which is born
spinning at $P=10$~ms.  If the unstable r-mode is present, the propeller
effect will not switch on until $\sim 200$~days.  If the r-mode is absent,
however, the propeller will come on after only $\sim 20$~days.  For a
$B=10^{12}$~G star born spinning at $P=1$~ms, the difference is even more
pronounced.  In this case the propeller will switch on after $\approx 3$~days
when there is no r-mode, as compared with $\sim 200$~days when the
unstable r-mode is present.  

\section{Concluding remarks}

We commenced this work with two prime objectives: to understand the impact
of supernova fallback accretion on the unstable r-modes of neutron stars
and any associated gravitational wave signal, and to investigate the combined
influence of accretion and r-mode instability on the spin rates of young
neutron stars.  Our model, simple as it is, has offered insight into both
questions.  Our results suggest that newly formed neutron stars accreting
fallback material will indeed experience a phase of perhaps several years 
during which the r-mode is
unstable.  This would have positive ramifications for detection of
gravitational waves from the unstable r-modes of neutron stars.  Moreover,
the model suggests that this is true even for stars that are born with low
spin rates.  The effect of magnetic field is however crucial.  While stars
with canonical ($\le 10^{13}$~G) field strengths are predicted to go r-mode
unstable, the strong propeller effect associated with magnetars appears to
preclude the r-mode from growing to the saturation level. The 
external torque simply spins the star down fast enough that
the r-mode is not given time to grow.
It is thus unlikely that we will see
a strong r-mode gravitational wave signal from a highly magnetized star.
The likelihood of obtaining a gravitational wave signal from an unstable
bar-mode is also low in our model. Spin-down caused by emission of
radiation from the r-modes is able to counteract even the strongest
accretion torque and prevent all but the stars with the most rapid initial
spins (and obviously those born spinning faster than the Kepler limit)
from going dynamically unstable to the bar-mode.  The propeller effect has
a similar preventive effect for the stars with the highest magnetic fields.

With regard to our second question, the model's predictions are consistent
with spin rates inferred from observations of young pulsars.  A canonical
neutron star is spun down within a few years by the combined action of fallback and r-mode to
a period similar to that inferred for the Crab pulsar at its birth.
Neutron stars with stronger magnetic fields, for which the propeller effect
is the dominant spin-down mechanism, achieve far longer periods in the
first year post-supernova.  Even for such highly magnetized stars, however, the effect of an
unstable r-mode on spin rate is significant. 

\section*{Acknowledgements}

We would like to thank Ian Jones, John Miller and Bryan Gaensler for helpful discussions. 
We also acknowledge support from the
 EU Programme
'Improving the Human Research Potential and the Socio-Economic
Knowledge Base', (Research Training Network Contract
HPRN-CT-2000-00137). A.W. acknowledges support from a PPARC postgraduate
studentship and N.A. acknowledges support from the Leverhulme
Trust in the form of a prize fellowship and PPARC grant PPA/G/1998/00606.

\section*{References}

\begin{flushleft}

Alpar M.A., 2001, ApJ, 554, 1245.

Andersson N., 1998, ApJ, 502, 708.

Andersson N., Kokkotas K.D., 2001, Int J Mod Phys
D, 10, 381.

Andersson N., Kokkotas K.D., Schutz B.F., 1999, ApJ, 510, 846.

Bethe H.A., Brown G.E., Lee C.-H., 2000, ApJ, 541, 918.

Bildsten L., Ushomirsky G., 2000, ApJ, 529, L33.

Brown E.F., Bildsten L., 1998, ApJ, 496, 915.

Brown G.E., Weingartner J.C., 1994, ApJ, 436, 843.

Burrows A., Lattimer J.M., 1986, ApJ, 307, 178.

Camilo F., Bell J.F., Manchester R.N, Lyne A.G., Possenti A., Kramer M.,
Kaspi V.M., Stairs I.H., D'Amico N., Hobbs G., Gotthelf E.V., Gaensler
B.M., 2001, ApJ, 557, L51.

Camilo F., Manchester R.N., Gaensler B.M., Lorimer D.R., Sarkassian J.,
2002, ApJ, 567. 

Chatterjee P., Hernquist L., Narayan R., 2000, ApJ, 534, 373.

Chevalier R.A., 1989, ApJ, 346, 847.

Colgate S.A., 1971, ApJ, 163, 221.

Colpi M., Geppert U., Page D., 2000, ApJ, 529, L29.

Cox D.P., Shelton R.L., Maciejewski W., Smith R.K., Plewa T., Pawl A.,
Rozyczka M., 1999, ApJ, 524, 179.

Crawford F., Gaensler B.M., Kaspi V.M., Manchester R.N., Camilo F., Lyne
A.G., Pivovaroff M.J., 2001, ApJ, 554, 152

Duncan R.C., Thompson C., 1992, ApJ, 392, L9.

Foster R.S., Lyne A.G., Shemar S.L., Backer D.C., 1994, AJ, 108, 175.

Friedman J.L., Morsink S.M., 1998, ApJ, 502, 714.

Gaensler B.M., Frail D.A., 2000, Nat, 406, 158.

Ghosh P., Lamb F.K., 1978, ApJ, 223, L83.

Gotthelf E.V., Vasisht G., Boylan-Kolchin M., Torii K., 2000, ApJ, 542, L37.

Haensel P., 1997, in Marck J.A., Lasota J.P., eds, Relativistic
Gravitation and Gravitational Radiation, Cambridge University Press, p.129.

Ho W.C.G., Lai D., 2000, ApJ, 543, 386.

Houck J.C., Chevalier R.A., 1991, ApJ, 376, 234.

Houser J.L., Centrella J.M., Smith S.C., 1994, Phys Rev Lett, 73, 1314.

Igumenshchev I.V., Narayan R., 2002, ApJ, 566, 137.

Illarionov A.F., Sunyaev R.A., 1975, A\&A, 39, 185.

Jones P.B., 2001, Phys Rev Lett, 86, 1384.

Kaspi V.M., 2000, in Wex N., Kramer M., Wielebinski R., eds, ASP Conf. Ser. Vol. 202, Pulsar
Astronomy - 2000 and beyond, p.405.

Kaspi V.M., Manchester R.N., Siegman B., Johnston S., Lyne A.G., 1994, ApJ,
422, L83.

Lattimer J.M., van Riper K.A., Prakash M., Prakash M.,
1994, ApJ, 425, 802. 

Lindblom L., Mendell G., 2000, Phys Rev D, 61, 104003.

Lindblom L., Owen B.J., 2002, astro-ph/0110558.

Lindblom L., Owen B.J., Ushomirsky G., 2000, Phys Rev D, 62, 084030.

Lindblom L., Tohline J.E., Vallisneri M., 2001, Phys Rev Lett, 86, 1152.

Lockitch K.L., Andersson N., Friedman J.L., 2001, Phys Rev D, 63, 024019.

Lyne A.G., Pritchard R.S., Graham-Smith F., Camilo F., 1996, Nat, 381, 497.

Marsden D., Lingenfelter R.E., Rothschild R.E., 2001, ApJ, 547, L45.

Menou K., Perna R., Hernquist L., 2001, ApJ, 554, L63.

Menou K., Esin A.A., Narayan R., Garcia M.R., Lasota J.P., McClintock J.E.,
1999, ApJ, 520, 276.

Migliazzo J.M., Gaensler B.M., Backer D.C., Stappers B.W., van der Swaluw
E., Strom R.G., 2002, astro-ph/0202063.

Mineshige S., Nomura H., Hirose M., Nomoto K., Suzuki T., 1997, ApJ, 489,
227. 

Murray S.S., Slane P.O., Seward F.D., Ransom S.M., Gaensler B.M., 2001,
astro-ph/0108489. 

Narayan R., Yi I., 1994, ApJ, 428, L13.

Narayan R., Yi I., 1995, ApJ, 444, 231.

Owen B.J., Lindblom L., Cutler C., Schutz B.F., Vecchio A., Andersson N.,
1998, Phys Rev D, 58, 084020.

Page,D. Prakash M., Lattimer J.M., Steiner A.W., 2000, 
Phys Rev Lett, 85, 2048.

Perna R., Hernquist L., Narayan R., 2000, ApJ, 541, 344.

Reynolds S.P., 1985, ApJ, 291, 152.

Rezzolla L., Lamb F.K., Shapiro S.L., 2000, ApJ, 531, L139.

Rothschild R.E., Lingenfelter R.E., Marsden D., 2001, astro-ph/0112121.

Shapiro S.L., Teukolsky S.A., 1983, Black Holes White Dwarfs and Neutron
Stars, Wiley.

Spruit H.C., 1999, A\&A, 341, L1.

Spruit H., Phinney E.S., 1998, Nat, 393, 139.

Stergioulas N., Font J.A., 2001, Phys Rev Lett, 86, 1148.

Torii K., Tsunemi H., Dotani T., Mitsuda K., Kawai N., Kinugasa K., Saito
Y., Shibata S., 1999, ApJ, 523, L69.

Zhang W., Marshall F.E., Gotthelf E.V., Middleditch J., Wang Q.D., 2001, ApJ,
554, L177.

\end{flushleft}
\onecolumn
\begin{table}
\caption[Pulsar/supernova remnant associations]{Young pulsars (PSR) that
have been identified with known supernova remnants (SNR).  Where the
braking index is not provided by observation, it  
is assumed to be similar to that of the Crab ($n= 2.5$).  The lower part of
the table illustrates the difficulty of estimating $t_{SNR}$ .

Notes:
(a) Measurement technique for $n$ is somewhat unorthodox. (b) Age
estimate made using proper motion of pulsar. (c) Using $t_{SNR} =
900$~years - for larger $t_{SNR}$ it is not possible to estimate $P_0$ using equation (29). (d)  As $t_{SNR}=t_c$ it
is not possible to estimate a value of $P_0$ using equation (29) - we can
only say that $P_0$ is small. (e) Age of
supernova remnant too uncertain to make a reliable estimate of $P_0$.  (f)
Estimated age of remnant much larger than $t_c$ making estimate of $P_0$
unfeasible.

Most references from
Kaspi (2000), additional data from: B0833-45 (Lyne et al 1996), B0540-69 (Zhang et al 2001, Reynolds
1985), PSR 1951+32 (Foster et al 1994,
Migliazzo et al 2002), J1811-1926 (Torii et al 1999),  J1846-0258 (Gotthelf
et al 2000), J205+6449 (Murray et al 2001), J1124-5916 (Camilo et al 2002), B1853+01 (Cox et al
1999),  B1509-58 (Kaspi et al 1994), J1119-6127 (Crawford et al
2001), J1016-5857 (Camilo et al 2001), B1757-24 (Gaensler \& Frail 2000).}

\label{pulsars}
\begin{tabular}{l|cccc|ccc}
PSR  & $P$ (ms) &  $n$ & SNR &  SNR age (kyr) &
$t_c$ (kyr) & $P_0$ (ms) &  B ($10^{12}$~G)\\
\hline
B0531+21 & 33 &  2.52  & Crab  & 0.95 (SN1054) & 1.2 & 18 &  3.8 \\
B0833-45   & 89  & 1.4 (a)  & Vela  &  9--30 & 11 &  13--57  &  3.4\\
B0540-69 & 50 &  1.81 & N158A & 0.8-1.1 & 1.7 & 34--39  &  5.0 \\
PSR 1951+32 & 39.5 &  & CTB 80  & 38-74 (b) & 107 & 21--33 &  0.49 \\
J1811-1926  & 65 &  & G11.2-0.3 & 1.7 (SN386) & 24  & 63 &  1.7 \\
J1846-0258 & 324 &  &  Kes 75 & 0.9-4.3 & 0.7 & 315 (c)  & 49 \\
J205+6449 & 66 & & 3C58 & 0.82 (SN1181) & 5.4 & 60 & 3.6 \\
J1124-5916 & 135 & & G292.0+1.8 & 1.7 & 2.9 & 90 & 10 \\
\hline
B1853+01 & 267 &  & W44 & 20 (d) & 20 & $<<P$ & 7.6 \\
B1509-58 &151 &  2.83 & MSH15-52 & (e) & 1.6 & (e)  & 15 \\
J1119-6127 & 408 & 2.91 & G292-05 & (e) & 1.6 &(e)  & 41 \\
J1016-5857 & 107 & & G284.3-1.8 & $\sim$ 10 (e) & 21 &(e)  & 3 \\
B1757-24 & 125 & & G5.4-1.2 & 39-170 (f) & 15.5 & (f) & 4 \\
\end{tabular}
\end{table}
\end{document}